%% file: prl_KKpipi_Y2175_GlueX_Final_SubmitArXive.tex
 \documentclass[
reprint,
superscriptaddress,
nofootinbib,
 amsmath,amssymb,
 aps,
 prl,
twocolumn,
]{revtex4-2}
\usepackage{graphicx}  
\usepackage{dcolumn}   
\usepackage{bm}        
\usepackage{amssymb}   
\usepackage{amsmath}   
\usepackage[mathlines]{lineno}
\usepackage{academicons}
\usepackage{orcidlink}

\begin{document}
\widetext
\title{Search for the $Y(2175)$ in the photoproduction cross section \\measurement of $\gamma p \to \phi \pi^{+} \pi^{-} p$ at GlueX}
\input{authors_JS_Oct3rd_FN.tex}
\date{\today}
\begin{abstract}
Based on 334\,pb$^{-1}$ of photoproduction data collected with the GlueX detector at Jefferson Lab,
we have measured for the first time the cross section of the exclusive reaction $\gamma + p \to \phi(1020) \pi^+\pi^- p$ by reconstructing the final state $K^+K^ -\pi^ +\pi^- p$ produced with a photon beam of energies between $8.0\,{\rm GeV}$ and $11.6\,{\rm GeV}$. 
Based on the measured differential cross section, we have performed a search for the strangeonium-like exotic candidate $Y(2175)$, recently renamed to $\phi(2170)$. This state has been reported by different $e^+e^-$ annihilation experiments and it is addressed here for the first time in a photoproduction experiment. We do not find evidence for this state when using the resonance parameters quoted by the Particle Data Group and provide upper limits on the photoproduction cross section. Instead, we find a structure at a mass of $m(\phi\pi^+\pi^-)= 2.24$\,GeV/$c^2$ with a statistical significance of about 5\,$\sigma$. The parameters of this structure differ from those quoted by the Particle Data Group for the $\phi(2170)$ and are consistent with a previous observation in $e^+e^-$ annihilation.
In addition, there is evidence for a second structure at 1.82\,GeV/$c^2$.
\end{abstract}

\maketitle
%
%
\paragraph{Introduction ---}
Since the beginning of the millennium, numerous unexpected states beyond the conventional quark model have been discovered, particularly in the charmonium region. These charmonium-like states are known as $XYZ$ states, and several are manifestly exotic, see {\it e.g.}~\cite{XYZ-review2021} for a recent review.

Searches have been conducted for analogous states within the light-quark sector. A new structure at 2.16\,GeV/$c^2$ with $J^{PC}=1^{--}$, originally denoted as $Y(2175)$, and meanwhile renamed to $\phi(2170)$ by the Particle Data Group (PDG)~\cite{PDG2022}, was initially reported by the BaBar collaboration in the initial-state radiation (ISR) process $e^+e^- \to \gamma_{\rm ISR} \phi f_0(980)$~\cite{BaBar2006,BaBar2012}. It was later confirmed by additional $e^+e^-$ collider experiments, including BES~\cite{Bes2008}, Belle~\cite{Belle2009}, and BESIII~\cite{BesIII2015}. The experimentally observed resonance parameters are largely consistent with each other within uncertainties. At BESIII, the decay channel to $K^+K^-$ has been studied recently~\cite{BesIII2019}, and a structure has been observed as well, however, with a slightly higher mass of about 2.24\,GeV/$c^2$. 

These observations have stimulated theoretical speculations about the nature of the $\phi(2170)$, especially as a potential strange partner of the famous charmonium-like $Y(4260)$ state~\cite{BaBar2005}, renamed to $\psi(4230)$ by the PDG~\cite{PDG2022}. Proposed interpretations of the internal structure include an $s\bar{s}g$ hybrid state~\cite{GJDingMLYan:2007}, an $(s\bar{s}s\bar{s})$ tetraquark state~\cite{ZGWang:2007,HXChen_etal:2008}, $\Lambda\bar{\Lambda}$ baryonium~\cite{EKlemptAZaitsev:2007}, a $\phi f_0(980)$ resonance~\cite{AMartinezTorresEtAl:2008}, and conventional $3^3S_1$~\cite{TBarnesNBlackPRPage:2003} and $2^3D_1$~\cite{XWangEtAl:2012,GJDingMLYan:2007} $\phi$ states.

The available experimental information for the $\phi(2170)$ is, however, restricted to $e^+e^-$ experiments. Information on the production in other processes, such as photoproduction, is highly desirable to sort out the proposed interpretations of the nature of this state.

According to the theoretical calculation in~\cite{TheoPredictPhotoY2175}, the $\phi(2170)$ is predicted to be produced by an 8\,GeV photon beam  with an estimated cross section of about 1\,nb (integral of the dashed blue curve in Fig.\,2 of~\cite{TheoPredictPhotoY2175}). Instead of focussing on the $\eta\phi$ decay channel as in that study, we have performed a measurement of the differential cross section ${\rm d}\sigma(\gamma p \to \phi\pi\pi p)/{\rm d}m$ to better compare with previous observations and benefit from a branching fraction expected to be similar or larger~\cite{BRphif0TophiEta}. Preliminary results of the current work were presented at the HADRON 2023~\cite{KG-FN_Hadron2023} and MESON 2023~\cite{KG-FN_Meson2023} conferences.
%
%
%
\paragraph{The Experiment ---}
The GlueX experiment is dedicated to mapping the light-meson spectrum and exotic states in photoproduction. It is operated at Jefferson Lab in Hall D and utilizes a linearly polarized photon beam impinging on a liquid-hydrogen target. The photon beam is produced via coherent bremsstrahlung of the electron beam, which is provided by the Continuous Electron Beam Accelerator Facility (CEBAF), on a 50 $\mu$m thick diamond wafer, with maximum polarization occurring in the coherent peak at about 9 GeV. The fixed-target spectrometer is a nearly hermetic detector optimized for reconstructing a wide range of charged as well as neutral final-state particles. Charged particle identification (PID) is achieved through measuring the time-of-flight and the energy loss (d$E$/d$x$) in the drift chambers. For detailed information on both the beamline and the detector, see~\cite{GlueXBeamlineDetector} and references therein.

The GlueX Phase-I data analyzed here were collected in three periods during 2017 and 2018 with a total integrated luminosity of 334\,pb$^{-1}$ for $E_\gamma > 8$\,GeV. To determine the reconstruction efficiencies for the studied reaction and the resolution of kinematic variables, Monte Carlo (MC) simulated event samples were generated using a phase-space generator. The GlueX detector response was simulated using GEANT4~\cite{MCsoftware_Geant4} and uncorrelated detector noise was extracted from randomly triggered real data and merged into the simulated events. The MC events are generated without any intermediate resonances to ensure uniform coverage along the three-body invariant mass $m(\phi\pi^+\pi^-)$ and to facilitate a reliable determination of the mass-dependence of the reconstruction efficiency.
The numbers of simulated events for the three periods is chosen to match the ratio of luminosities of the real data periods.

We select fully reconstructed events by applying the following criteria. Using the PID information, we form event candidates for the final-state particle hypothesis $(K^+ K^- \pi^+\pi^- p)$. A kinematic 4-constraint (4C) fit of the $(K^+ K^- \pi^+\pi^- p)$ final to the $(\gamma p)$ initial state enforces energy and momentum conservation and constrains the particles to a common production vertex. We restrict the beam photon energy to $E_\gamma > 8$\,GeV and, to reduce contributions from baryon excitations of the target proton, we require the squared four-momentum transfer $-t < 1$ GeV$^2/c^2$. We apply the same $t$ selection when generating the MC data. Thus, the cross section results presented are ``fiducial'' measurements restricted to $|t| < 1$ GeV$^2/c^2$. To further suppress background from the reaction $\gamma p \to \phi \pi^- \Delta^{++}$, we require $m(\pi^+p) > 1.35$\,GeV/$c^2$. For efficiency correction, we assume a phase-space distribution for the $m(\pi^+p)$ spectrum in the excluded region.
The exclusivity of the events is ensured by applying a cut on the missing mass squared of the total reaction, $|MM^2| < 50$\,MeV$^2$/$c^4$, and on the reduced $\chi^2$ of the kinematic fit, $\chi^2_{\rm 4C + vertex}/n_{\rm dof} < 70/11$.
%
%
%
\begin{figure}[tp!]
    \begin{center}
      \includegraphics[clip,trim= 4 2 0 10,width=0.48\linewidth, angle=0]{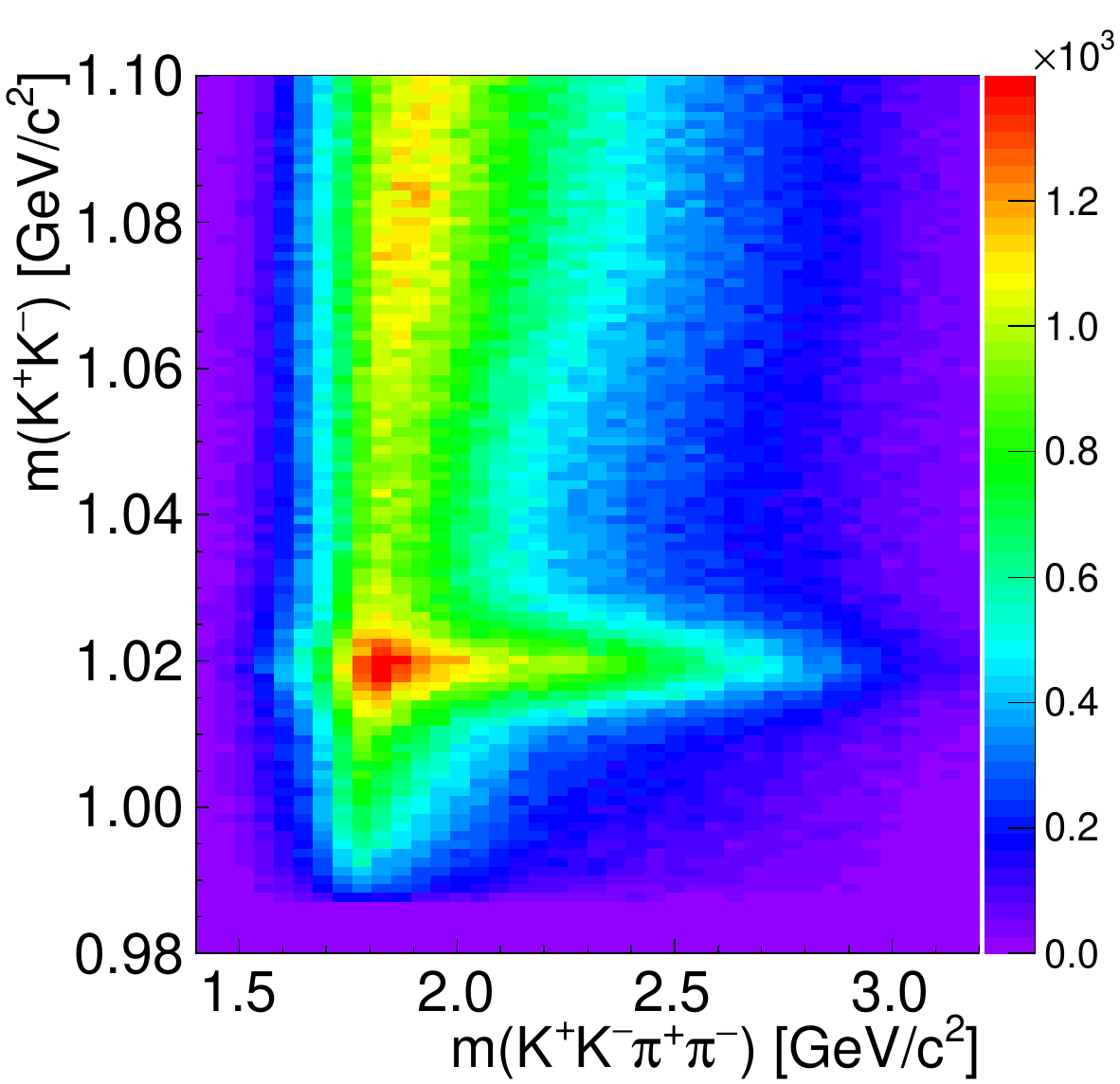}
     \includegraphics[clip,trim= 4 0 5 0,width=0.5\linewidth, angle=0]{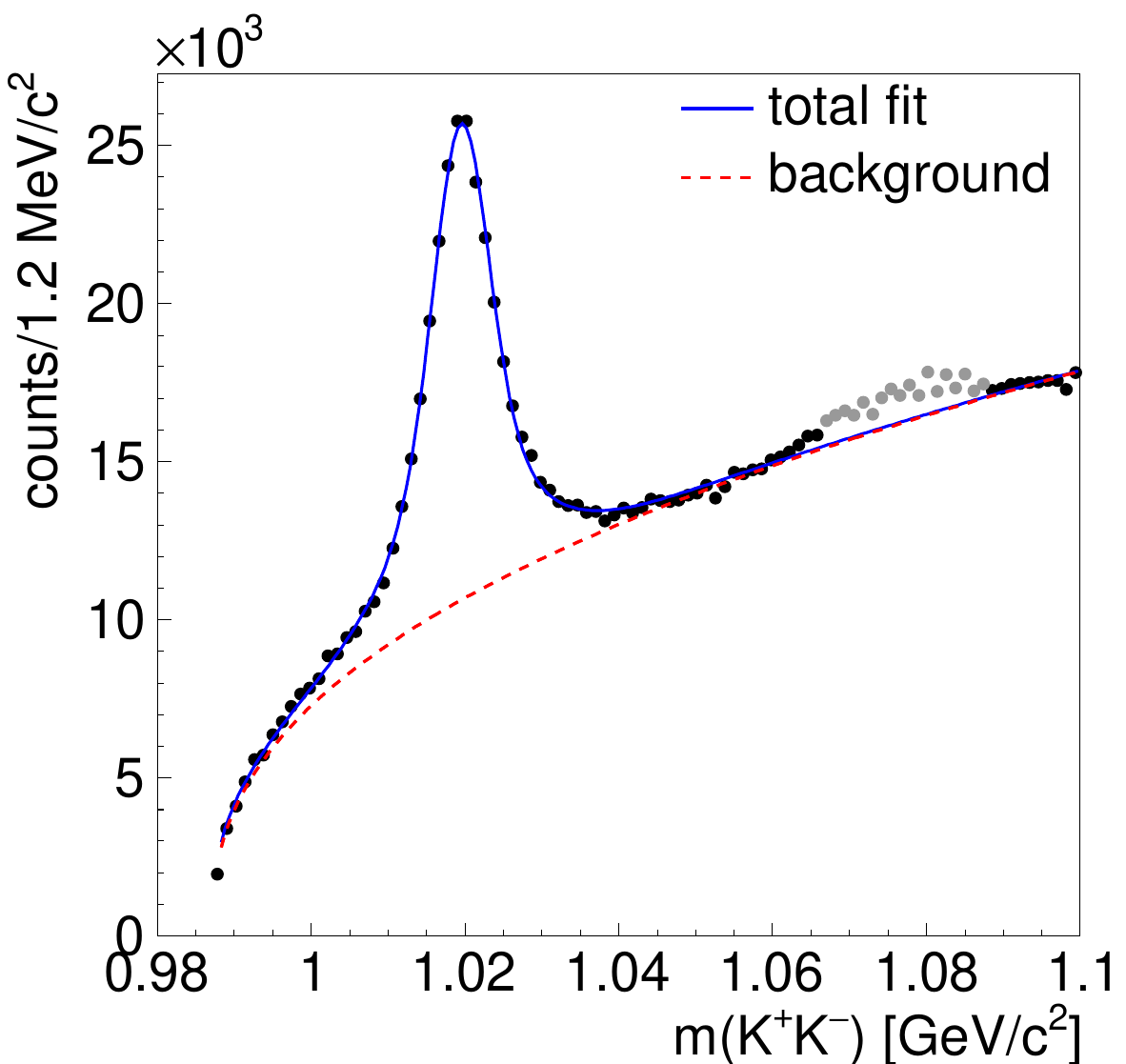}
   \end{center}
   \vspace{-0.9cm}
     \begin{picture}(1,1)
 \put(-72,92){(a)}
 \put(25,92){(b)}
  \end{picture}
    \caption{Observed $\phi(1020)$ signal. {(a)} Distribution of invariant masses $m(K^+K^-)$ vs. $m(K^+K^-\pi^+\pi^-)$.
      {(b)} $K^+K^-$ mass spectrum with fitted signal and background function.}
       \label{fig:Fig1}
      \vspace{-0.3cm}
\end{figure}
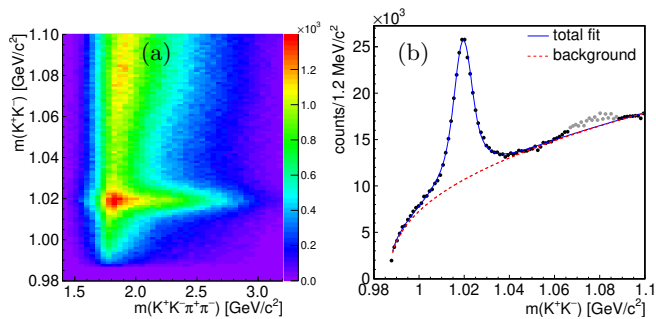
The latter two selection cuts are optimized using the significance $Z_{\phi}=S/\sqrt{S+B}$ of the $\phi(1020)$-signal yield $S$ with respect to the background yield $B$ in data.

In-depth investigations resulted in an appropriate fit model consisting of a sum of two Voigtians~\cite{voigtian} (DV) constrained to identical resonance parameters for the $\phi(1020)$ peak plus an empirical function to describe the background:
\newline
\vspace{0.1cm}
~$f_{\rm fit}(m) = DV (m; m_0, \Gamma_0, \sigma_1, \sigma_2, R) + \vert m - m_t \vert^p \, e^{-\lambda m}$,
\newline
for $m \equiv m(K^+K^-) > m_t$. Here, $(m_0,\Gamma_0)$ are the Breit-Wigner (BW) resonance parameters of the $\phi(1020)$, $\sigma_{1,2}$ represent the mass resolution parameters, and $R$ the relative contribution of the second Voigtian. In the background model, $m_t= 2 m_K$, where $m_K$ is the kaon mass, represents the kinematic threshold, and the parameters $p$ and $\lambda$ determine the background shape. The data points around $m(K^+K^-) = 1.08$\,GeV/$c^2$ colored in grey in Fig.~\ref{fig:Fig1}b are excluded from the fit to avoid bias arising from this region that is not well described by the background model.
The natural decay width of the $\phi$ is fixed to the current best value reported by the PDG~\cite{PDG2022}.
A clear $\phi$-band is visible in the $m(K^+K^-)$ vs. $m(K^+K^-\pi^+\pi^-)$ invariant mass distribution (Fig.~\ref{fig:Fig1}a).
The $\phi$-peak in data is well described by the fit function (Fig.~\ref{fig:Fig1}b).

We have measured the cross section as a function of the $\phi \pi^+\pi^-$ mass by subdividing the data into 40 equidistant 45\,MeV/$c^2$-wide $m(\phi\pi^+\pi^-)$ bins in the range from 1.4 to 3.2\,GeV/$c^2$.
For each bin at the $\phi\pi^+\pi^-$ mass $m_i$, we extract the $\phi\pi^+\pi^-$ event yield by a $\chi^{2}$-fit of the model $f_{\rm fit}(m)$ to the $m(K^+K^-)$ distribution for this mass bin via the $\phi$-yield given by the integral of the signal part of $f_{\rm fit}(m)$. In these fits, the parameters $m_0$, $\sigma_1$, $\sigma_2$ are fixed to the values obtained from fitting the overall $m(K^+K^-)$ spectrum (Fig.~\ref{fig:Fig1}b). 

\begin{figure*}[tph!]
  \begin{minipage}[h]{.48\textwidth}
    \begin{center}
      \vspace{-0.1cm}
      \includegraphics[clip,trim= 0 0 0 0,width=0.8\linewidth, angle=0]{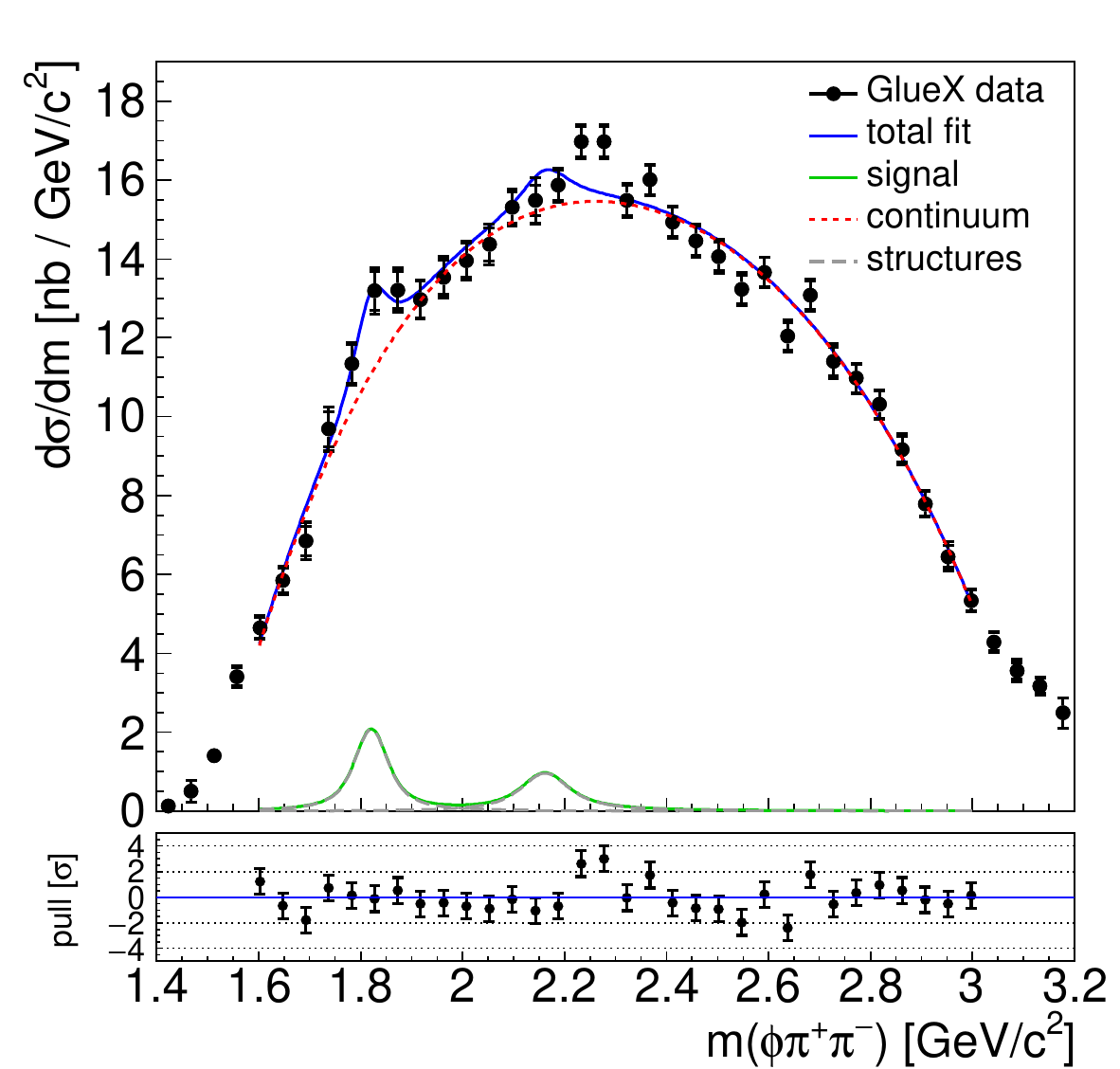}    
     \end{center}
  \end{minipage}
  \hfill
  \begin{minipage}[h]{.48\textwidth}
    \begin{center}
      \vspace{-0.3cm}
      \includegraphics[clip,trim= 0 0 0 0,width=0.8\linewidth, angle=0]{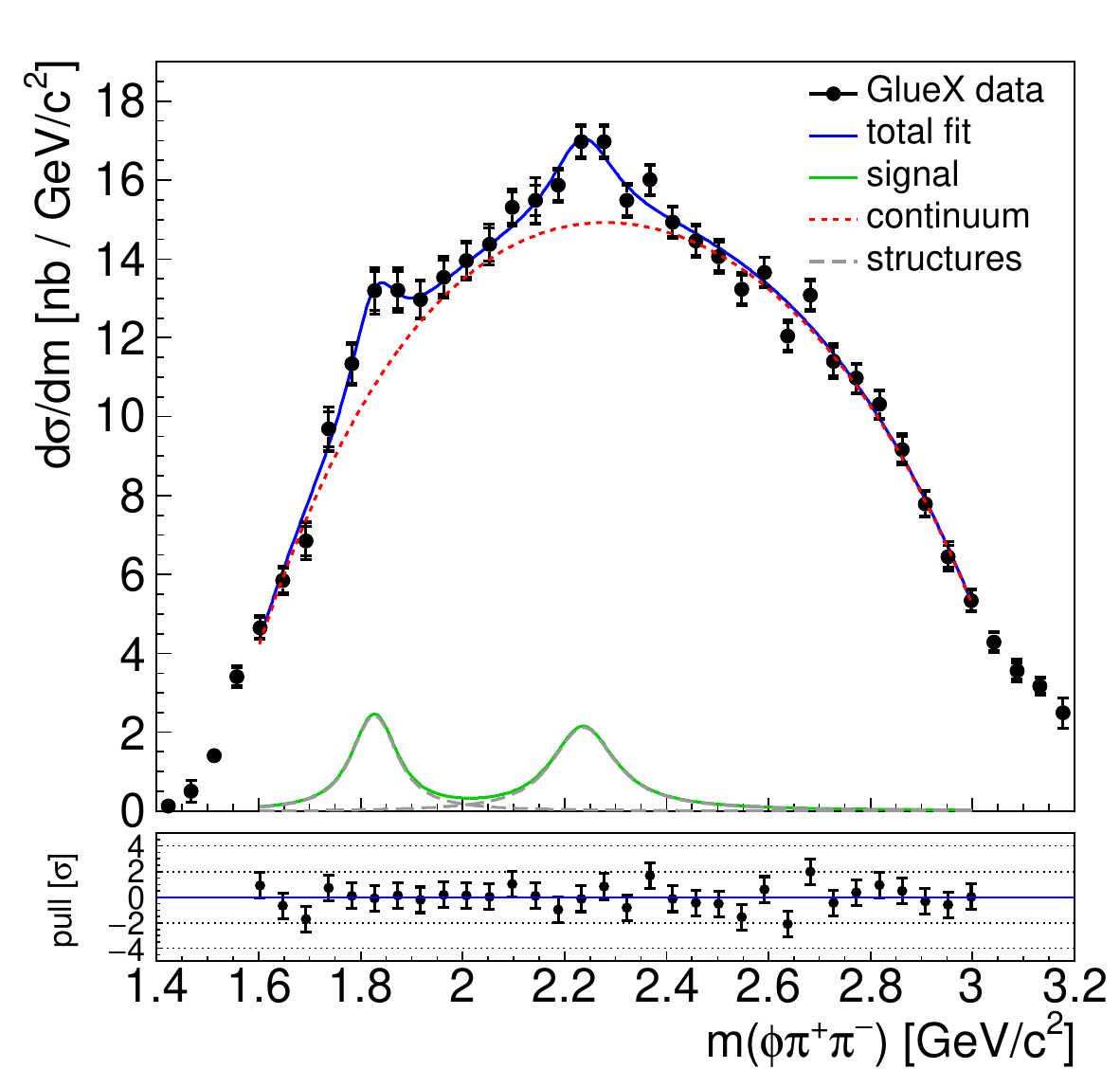}
    \end{center}
  \end{minipage}
  \begin{picture}(1,1)
 \put(-450,70){(a)}
 \put(-400,5){(Fit A)}
 \put(-188,70){(b)}
 \put(-138,5){(Fit B)}
  \end{picture}

  \caption{Measured differential cross section ${\rm d}\sigma/{\rm d}m$ for the reaction $\gamma p \to \phi(1020) \pi^+\pi^- p$ (data points) and search for the $\phi(2170)\to \phi(1020) \pi^+\pi^-$ in photoproduction. The data are modeled using Breit-Wigner functions to describe the peak-like structures at 1.8 and 2.2\,GeV/$c^2$ (grey dashed curves).
    {(a)} Fit A, where the Breit-Wigner parameters for the structure at 2.2\,GeV/$c^2$ are fixed to the PDG average values for the $\phi(2170)$~\cite{PDG2022}.
    {(b)} Fit B, where the Breit-Wigner parameters for the structure at 2.2\,GeV/$c^2$ are fixed to the values measured by BESIII~\cite{BesIII2019}.
  }
       \label{fig:CrossSectionsAndFits}
\vspace{-0.2cm}
\end{figure*}
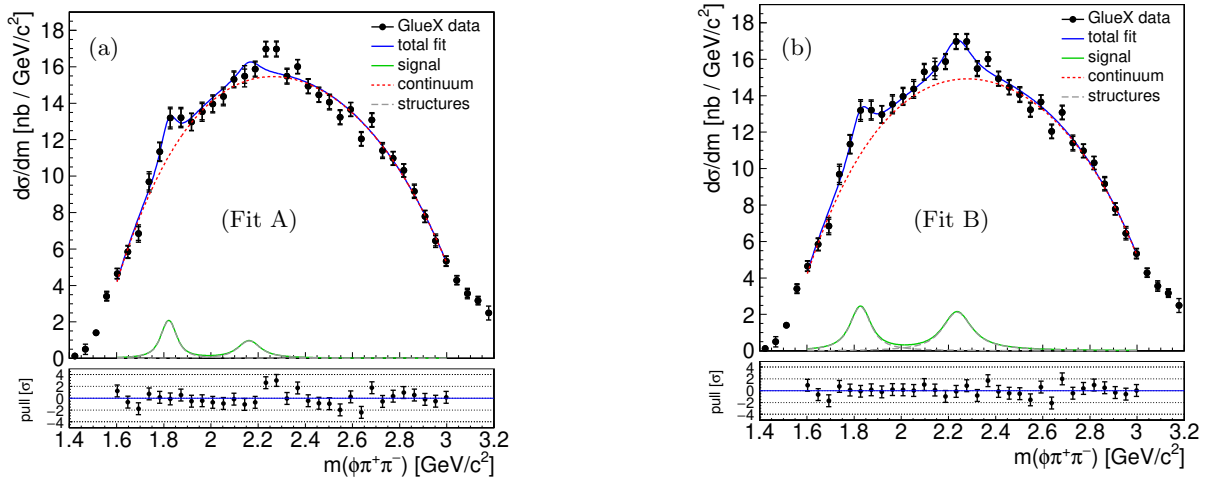

\paragraph{Differential cross section measurements ---}
The cross section for photoproduction of $\phi \pi^{+} \pi^{-}$ on the proton is computed for each $m(\phi \pi^+\pi^-)$ bin $m_i$ according to
\begin{equation}
  \label{XsecFormula}
\sigma(m_i) = \frac{N_\phi(m_i)}{\epsilon(m_i)\, F \, d_{\rm target} \, {\cal B}(\phi \to K^+K^-)}~,
\end{equation}
where $N_\phi(m_i)$ is the number of reconstructed $\phi$ mesons in each $m(\phi \pi^+\pi^-)$ mass bin $m_i$, $\epsilon(m_i)$ is the mass-dependent reconstruction efficiency determined bin-by-bin, $F$ is the tagged photon beam flux, $d_{\rm target} = 1.271 \,{\rm b}^{-1}$ is the target thickness, and ${\cal B}(\phi \to K^+ K^-) = (49.1 \pm 0.5)$\,\% is the branching fraction~\cite{PDG2022}.
The efficiencies $\epsilon(m_i)$ are obtained by dividing the number of reconstructed by the number of generated phase-space MC events for each bin $m_i$. The resulting values range from 4\,\% to 6\,\%.
Possible backgrounds were studied using events generated by PYTHIA~\cite{bggenPythia}, which were simulated and reconstructed using the same procedure as for the phase-space MC events. These studies show that the most dominant background originates from the non-resonant reaction $\gamma p \to  \pi^+ \pi^- K^+ K^- p$.
None of the studied background channels produce a peaking structure in the vicinity of the $\phi(1020)$.

Systematic uncertainties are summarized in Tab.\,\ref{tab_sys_FitB} in the Appendix.
We distinguish and treat uncorrelated and correlated systematic uncertainties separately. The former are independent from bin to bin, and are summed quadratically for each bin. Correlated uncertainties, including beam flux, branching ratio and track rconstruction, affect the cross section measurements in all bins in the same way.   

For each source of uncorrelated uncertainties affecting the differential cross section measurement ${\rm d}\sigma/{\rm d}m$ (Type\,I), the full measurement is repeated for the correspondingly varied parameter or selection criterion. The differences from the nominal result are summed up quadratically to form the total uncorrelated systematic uncertainty. These uncertainties are reflected in the error bars of the measured data points of the ${\rm d}\sigma/{\rm d}m$ spectra, while the correlated ones are not.

The measured differential $\phi\pi^+\pi^-$ photoproduction cross section ${\rm d}\sigma/{\rm d}m$ is shown as a function of the $m(\phi\pi^+\pi^-)$ mass in Fig.\,\ref{fig:CrossSectionsAndFits}.
The inner error bars of the data points represent the statistical and the outer ones the quadratic sum of the statistical and the uncorrelated systematic uncertainties.   
\paragraph{Search for the $\phi(2170)$ and other resonances ---}
Based on the measured differential $\phi\pi^+\pi^-$ cross sections, we perform a search for the $\phi(2170)$ and other resonances.

We observe two peak-like signals at $\phi\pi^+\pi^-$ masses of about 1.8 and 2.2\,GeV/$c^2$ on top of a featureless background. We model each peak by a relativistic BW shape and use a $4^{th}$-order Chebyshev polynomial to describe the background. For the peak at 2.2\,GeV/$c^2$, we use different sets of fixed resonance parameters for the $\phi(2170)$, whereas those for the peak at 1.8\,GeV/$c^2$ are kept free, when performing $\chi^2$ fits to the measured ${\rm d}\sigma/{\rm d}m$ distribution in the range from $m(\phi\pi^+\pi^-)=1.6$ to 3.0\,GeV/$c^2$.   
The $\phi\pi^+\pi^-$ mass resolution of about 16.3\,MeV/$c^2$ obtained from the MC simulated data is taken into account by convolving the fit function with a corresponding Gaussian distribution.

Based on the fit results, we determine $\sigma_{X}$, which is the measured production cross section of a resonance $X$ times the unknown decay branching fraction ${\cal B}(X\rightarrow \phi\pi^+\pi^-)$ for $|t| < 1$ GeV$^2/c^4$, by integrating the corresponding Breit-Wigner intensity of the fit model over the fit range. We also provide corresponding upper limits at the 90\,\% confidence level (CL90), statistical significances $Z$, and (where applicable) the determined resonance parameters $(m, \Gamma)$.

The statistical significances of the observed resonance signals are determined according to Wilks theorem, based on the change $\Delta\chi^{2}$ in the binned fits with and without the given resonance being included in the fit model, taking into account the change in the number of degrees of freedom $\Delta n_{\rm dof}$.  
We quote the statistical significances as $Z = Z_{\rm sys}(Z_{\rm stat})$, where $Z_{\rm stat}$ includes only the statistical uncertainties, while $Z_{\rm sys}$ also includes the uncorrelated systematic uncertainties of ${\rm d}\sigma/{\rm d}m$.
The correlated systematic uncertainties are not taken into account here, since they just scale all measured data points by a common factor, without any impact on the significances of the structures.

The systematic uncertainties for all fit results are treated as follows. We repeat the fits to the differential cross section ${\rm d}\sigma/{\rm d}m$ for the correspondingly varied parameters of all sources as listed in Tab.\,\ref{tab_sys_FitB}. The quadratic sum of the differences in the cross section between the nominal and varied parameters are taken as the systematic uncertainties. Whereas the Type\,I uncertainties affect the spectrum ${\rm d}\sigma/{\rm d}m$ itself and the nominal fit model is used, the Type\,II uncertainties affect the fit model applied to the nominal ${\rm d}\sigma/{\rm d}m$ spectrum (Fig.\,\ref{fig:CrossSectionsAndFits}). Type\,II sources are the $m(\phi\pi\pi)$ fit range, the resonance parameters $(m, \Gamma)$ and the background description. We vary the resonance parameters within $\pm 1\sigma$ of the quoted uncertainties and the degree of the Chebyshev polynomial for the background description from 4th to 5th order, and subsequently repeat the fits. We take the larger difference as systematic uncertainty in case of parameter variations.

We apply different fit models for our search. One model (Fit\,A) describes the peak at 2.2\,GeV/$c^2$ by a relativistic BW shape for the $\phi(2170)$ with spin $J$=$1$ and the fixed averaged values $(m_{\phi(2170)},\Gamma_{\phi(2170)}) = (2163 \pm 7 {\rm MeV}/c^2, 100^{+31}_{-21}\,{\rm MeV})$ quoted by the PDG~\cite{PDG2022}, and the peak at at 1.8\,GeV/$c^2$ by a second BW shape with free resonance parameters. 
While the structure at 1.8\,GeV/$c^2$ is described reasonably well, the overall fit quality especially around 2.2\,GeV/$c^2$ is rather poor as quantified by ${\chi}^2$/$n_{\rm dof}$ = 44.6/23 = 1.94 (Fig.\,\ref{fig:CrossSectionsAndFits}a).
The resultant parameters from the search for the $\phi(2170)$ resonance (Fit A) are summarized below.  
\begin{align*}
\begin{aligned}
 \sigma_{\phi(2170)} & =  142 \pm 57\,{\rm(stat.)} \pm 76 \,{\rm(sys.)} \pm 25 \,{\rm(corr.)}\mbox{ pb} \\
 \qquad Z &= 1.9\,(2.0)\sigma~;~~\sigma_{\phi(2170)}  <  272   \mbox{ pb (CL90)}
\end{aligned}
\end{align*}
Using the PDG averages for the $\phi(2170)$ parameters, we do not find evidence for this state in our data, and thus provide an upper limit on the photoproduction cross section for this state.
For the measured absolute production cross sections, the total correlated (bin-independent) uncertainty is quoted separately, and it is taken into account for the determination of the provided upper limits.
Given the poor overall fit quality we do not quote the fit parameters of the lower mass structure for this fit.

A second model (Fit\,B) uses instead of the PDG averaged values for the $\phi(2170)$ the resonance parameters $(m_{Y(2240)},\Gamma_{Y(2240)}) = (2239.2 \pm 13.4\,{\rm MeV}/c^2, 139.8\pm 24.0\,{\rm MeV})$ as measured by BESIII~\cite{BesIII2019}. Though this particular observation is quoted by the PDG in the context of the $\phi(2170)$, the PDG does not include it in their average. With ${\chi}^2$/$n_{\rm dof}$ = 24.7/23 = 1.07, the quality of Fit\,B is significantly higher than that of Fit A, mainly due to the highly improved description of the observed structure at about 2.2\,GeV/$c^2$ (Fig.\,\ref{fig:CrossSectionsAndFits}b). Remarkably, the shape of this enhancement matches exceptionally well with the ones reported by BESIII and BaBar in the different production process $e^+e^-\to K^+K^-$~\cite{BesIII2019,BaBar2013}.

The resultant values for the observed $Y(2240)$ obtained from the second model (Fit B) are shown below, including the determined resonance parameters $(m, \Gamma)$ for the structure at about 1.8 GeV/$c^2$.
\begin{align*}
\begin{aligned}
 \sigma_{Y(2240)} & =  475 \pm 70\,{\rm(stat.)} \pm 93 \,{\rm(sys.)} \pm 84 \,{\rm(corr.)}\mbox{ pb} \\
 \qquad Z &= 4.8\,( 5.5)\sigma~;~~\sigma_{Y(2240)}  <  658   \mbox{ pb (CL90)}\\
 \sigma_{X(1830)}  & = 373 \pm 108\,{\rm(stat.)} \pm 90 \,{\rm(sys.)} \pm 66  \,{\rm(corr.)} \mbox{ pb} \\
 \qquad Z &= 3.6 \; (3.7)\sigma~;~~\sigma_{X(1830)}   < 571 \mbox{ pb (CL90)} \\
m_{X(1830)}       & = 1828 \pm 12\,{\rm(stat.)} \pm 15 \,{\rm(sys.)} \mbox{ MeV/}c^2 \\ 
\Gamma_{X(1830)}  & = 94 \pm 33\,{\rm(stat.)} \pm 37 \,{\rm(sys.)} \mbox{ MeV} \\[1ex]
\end{aligned}
\end{align*}
The $X(1830)$ parameters obtained from this fit are consistent with those listed by the PDG for the $\phi_3(1850)$. However, the $\phi_3(1850)$ has never been observed in photoproduction nor to decay to $\phi\pi\pi$.

Furthermore, the $X(1830)$ lies in the vicinity of the $K^*(892)^0\bar{K^*}(892)^0$ threshold, and we observe a significant contribution in the data from the topology $\gamma p \rightarrow K^*(892)\bar{K}^*(892) p \rightarrow K^+K^-\pi^+\pi^- p$. Since the cross section is measured by fitting the $m(K^+K^-)$ spectra to obtain the $\phi$-yield, there is no direct contribution from this topology (as shown in the Supplemental Material in the Appendix).

The $X(1830)$ could, however, also be generated through the interaction between a $K^*(892)\bar{K}^*(892)$ pair and subsequent rescattering of the outgoing $K^+K^-$ pair to form a $\phi(1020)$, which may be enhanced due to the limited phase space available near the threshold. The observed mass of the $X(1830)$ is also consistent with theoretical predictions of a dynamically-generated resonance or a molecular bound $K^*\bar{K^*}$ state~\cite{KstarKstarMolecule_a,KstarKstarMolecule_b}.

When leaving the resonance parameters free for both structures in our fit model (Fit C), the obtained $Y(2240)$ parameters are within uncertainties consistent with those from Ref.~\cite{BesIII2019} and the $X(1830)$ parameters with those obtained from Fit B.
The obtained parameter values are
\begin{align*}
\begin{aligned}
m_{Y(2240)}       & = 2249 \pm 12\,{\rm(stat.)} \pm 10 \,{\rm(sys.)} \mbox{ MeV/}c^2 \\ 
\Gamma_{Y(2240)}  & = 153 \pm 54\,{\rm(stat.)} \pm 76 \,{\rm(sys.)} \mbox{ MeV/}c^2 \\[1ex]
m_{X(1830)}       & = 1828 \pm 12\,{\rm(stat.)} \pm 7 \,{\rm(sys.)} \mbox{ MeV/}c^2 \\ 
\Gamma_{X(1830)}  & = 95 \pm 34\,{\rm(stat.)} \pm 27 \,{\rm(sys.)} \mbox{ MeV}. \\[1ex]
\end{aligned}
\end{align*}

Our result supports the possible hypothesis that the enhancement around 2.24\,GeV/$c^2$ observed by BESIII, BaBar, and reported here might not correspond to the $\phi(2170)$ but could instead be a distinct resonance.

A full partial-wave analysis could further disentangle this 5-body final state,
which is beyond the scope of this letter and might be addressed based on a larger data set in the future.
\paragraph{Summary ---}
A first measurement of the differential photoproduction cross section for the reaction $\gamma p \to \phi \pi^{+} \pi^{-} p$ has been performed with the GlueX experiment. Based on the measured cross section, a search for the $\phi(2170)$ and further resonances has been performed.

We do not find evidence for $\phi(2170)$ with the averaged resonance parameters quoted by the PDG and thus provide upper limits on the photoproduction cross section for this state. We observe a structure at a mass of $m(\phi\pi^+\pi^-)= 2.24$\,GeV/$c^2$ with a statistical significance of about 5\,$\sigma$ that we find consistent with previous observations in $e^+e^-$ annihilation~\cite{BesIII2019,BaBar2013}. The statistical significance of this state is $5.5\,\sigma$ considering only statistical uncertainties and is reduced to $4.8\,\sigma$ when systematic uncertainties are included. 
The measured signal strength of the $Y(2240)$ is consistent with theoretical predictions for photoproduction of the $\phi(2170)$~\cite{TheoPredictPhotoY2175}. For completeness, we provide upper limits on the photoproduction cross section also for the $Y(2240)$.

We find evidence for a second structure at a mass of $m(\phi\pi^+\pi^-)= 1.82$\,GeV/$c^2$ with a significance of more than $3\,\sigma$ that substantially improves the description of the data when included in the fit model.
This structure is roughly consistent with the $\phi_3(1850)$ in the yet unobserved decay to $\phi\pi^+\pi^-$. However, this $m(\phi\pi^+\pi^-)$-region is also sensitive to $K\bar{K}$ rescattering near the $K^*\bar{K}^*$ threshold, which could be responsible for this structure in the cross section.
\\
\noindent
\input acknowledgement.tex   

\clearpage
\subsection{Appendix}
\paragraph{Systematic uncertainties  ---} 
The sources of uncorrelated systematic uncertainties are listed and summarised together with values for our main fit (Fit B) in Tab.\,\ref{tab_sys_FitB}.  
In addition, three sources for correlated systematic uncertainties are considered: (i) the uncertainty of the measured photon beam flux (5\,\%), (ii) the systematic uncertainty of track reconstruction (3\,\% for $\pi^\pm/K^\pm$ and 5\,\% for $p$, and thus in total $4\cdot3 +5 = 17$\,\%), and (iii) the uncertainty quoted by the PDG on the measured branching ratio ${\cal B}(\phi \to K^+K^-)$ (1\,\%). Adding them in quadrature yields a total correlated systematic uncertainty $d\sigma_{\rm corr} = 17.8\,\%$.
As discussed above, $d\sigma_{\rm corr}$ is not included in the total error bars of the data points of the differential cross section spectra (Fig.\,\ref{fig:CrossSectionsAndFits}).
\\
\\
\begin{table}[bph!]
\centering
\caption{Sources of uncorrelated (Type\,I) and fit-model related (Type\,II) systematic uncertainties, and the resulting relative uncertainties $d\sigma_{Y(2240)}$ for `Fit B' ({\it i.e.} the fit with two resonances, one fixed to the parameters from \cite{BesIII2019}).
  Three further sources of correlated uncertainties are considered in addition as described in the Appendix.
}
\label{tab_sys_FitB}
\begin{tabular}{l|l|r} 
  & Source of uncertainty    & $\Delta\sigma_{Y(2240)}$ [\%]   \\ \hline
  & Variation of cut on $\chi^2$              &   6.8   \\ 
  & Variation of cut on $|MM^2|$              &   2.2   \\ 
  & Beam photon bckgrd. subtraction         &   2.8   \\ 
  & $m(K^+K^-)$ binning             &  0.1    \\ 
  & $m(K^+K^-)$ fit range for $\phi$-yields              &  1.8    \\ 
  & $\phi$-peak position floating           &  0.0    \\ 
  & Veto $\phi$-signal fit around 1.08\,GeV/$c^2$           &   0.8   \\ 
  & $\phi$ fit model MC             &   1.2   \\ 
  & Merging of the three data sets                    &   0.8   \\ \hline
  & $m(\phi\pi^+\pi^-)$ fit range                       &   1.6   \\ 
  & Background parametrization             &   0.6   \\ 
  & Breit-Wigner model (rel./non-rel.)             &   3.5   \\ 
  & Variation of $m_{Y(2240)}$        &   3.1   \\ 
  & Variation of $\Gamma_{Y(2240)}$       &  17.0   \\ 
  & Variation of spin in rel. BW                  &   1.5   \\ \hline
  & Total uncorr. systematic uncertainty                &  19.5   \\ \hline
  & Statistical uncertainty         &  14.8   \\ \hline
  & Total uncertainty without correlation                &  24.5   \\ \hline
\end{tabular}
 \begin{picture}(1,1)
 \put(-225,25){\rotatebox{90}{Type I}}
 \put(-225,-50){\rotatebox{90}{Type II}}
  \end{picture}
\end{table}
~~~~~~~~~~~~~~~~~~~~~~~~~~~~~~~~~~~~~~~~~~~~~~~~~~~~~~~~~~~~~~~~~~~~~~~~~~~~~~~~~~~~~~~~~~~~~~~~~~~~~~~~~~~~~~~~~~~~~~~~~~~~~~~~~~~~~~~~~~~~~~~~~~~~~~~~~~~~~~~~~~~~~~~~~~~~~~~~~~~~~~~~~~~~~~~~~~~~~~~~~~~~~~~~~~~~~~~~~~~~~~~~~~~~~~~~~~~~~~~~~~~~~~~~~~~~~~~~~~~~~~~~~~~~~~~~~~~~~~~~~~
\clearpage
\begin{figure*}[tph!]
      \vspace{-0.3cm}
  \begin{minipage}[h]{.32\textwidth}
    \begin{center}
      \vspace{-0.1cm}
      \includegraphics[clip,trim= 0 0 0 0,width=1.0\linewidth, angle=0]{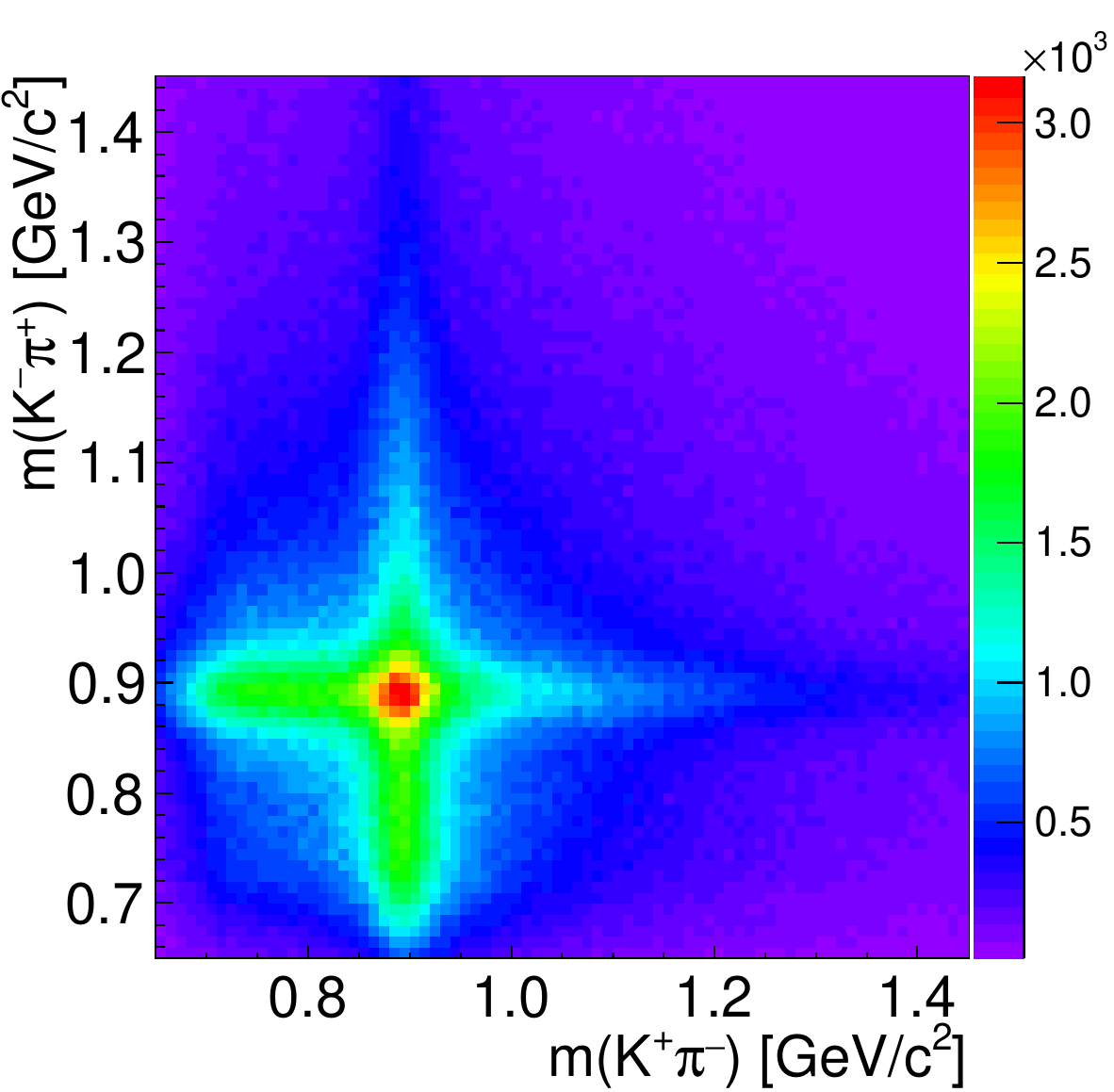}    
     \end{center}
  \end{minipage}
  \hfill
  \begin{minipage}[h]{.32\textwidth}
    \begin{center}
      \vspace{-0.1cm}
      \includegraphics[clip,trim= 0 0 0 0,width=1.0\linewidth, angle=0]{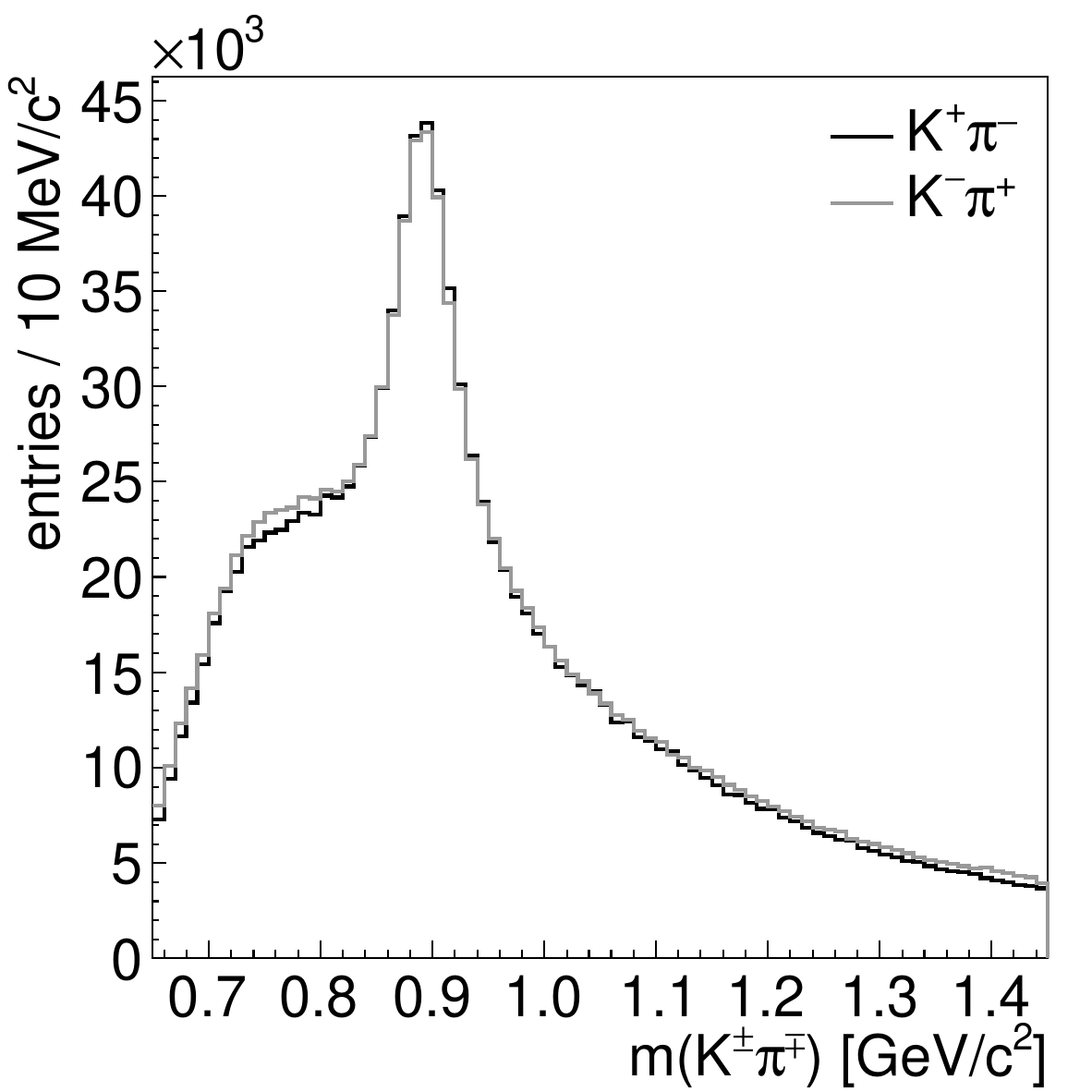}
    \end{center}
  \end{minipage}
  \hfill
  \begin{minipage}[h]{.32\textwidth}
    \begin{center}
      \vspace{-0.1cm}
      \includegraphics[clip,trim= 0 0 0 0,width=1.0\linewidth, angle=0]{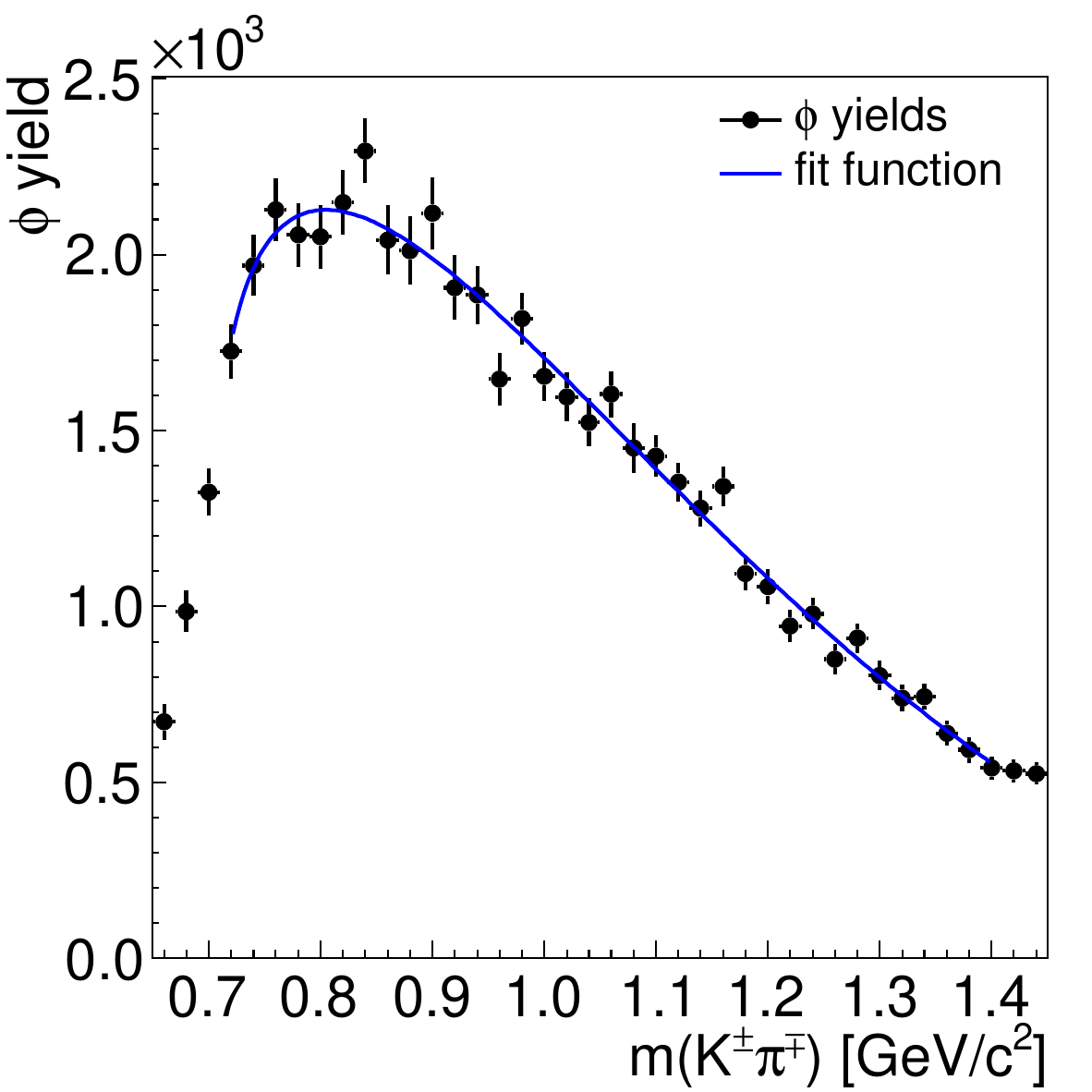}
    \end{center}
  \end{minipage}
  \begin{picture}(1,1)
    \put(-475,59){(a)}
    \put(-308,59){(b)}
    \put(-138,59){(c)}
  \end{picture}
      \vspace{-0.3cm}
  \caption{Further supplemental studies concerning the additional structure at about 1.8\,GeV/$c^2$.
    {(a)} Correlated $m(K^-\pi^+)$ vs. $m(K^+\pi^-)$ invariant mass after all event selection cuts applied, showing a strong $K^{*0}\bar{K}^{*0}$ contribution in the data.
    {(b)} Projections of the $m(K\pi)$ invariant mass distributions after requiring $|m(K^\pm\pi^\mp) - 892\,{\rm MeV/}c^2| < 100$\,MeV/$c^2$.
    {(c)} Illustration of the insensitivity of the $\phi$-yield extraction method to backgrounds from $K^{*0}\bar{K}^{*0}$ events: No  $K^*(892)^0$ signal is visible in the combined extracted $\phi$-yield as a function of $m(K^\pm\pi^\mp)$.
  }
  \label{fig:SupplementaryMaterial}
 \vspace{-0.3cm}
\end{figure*}
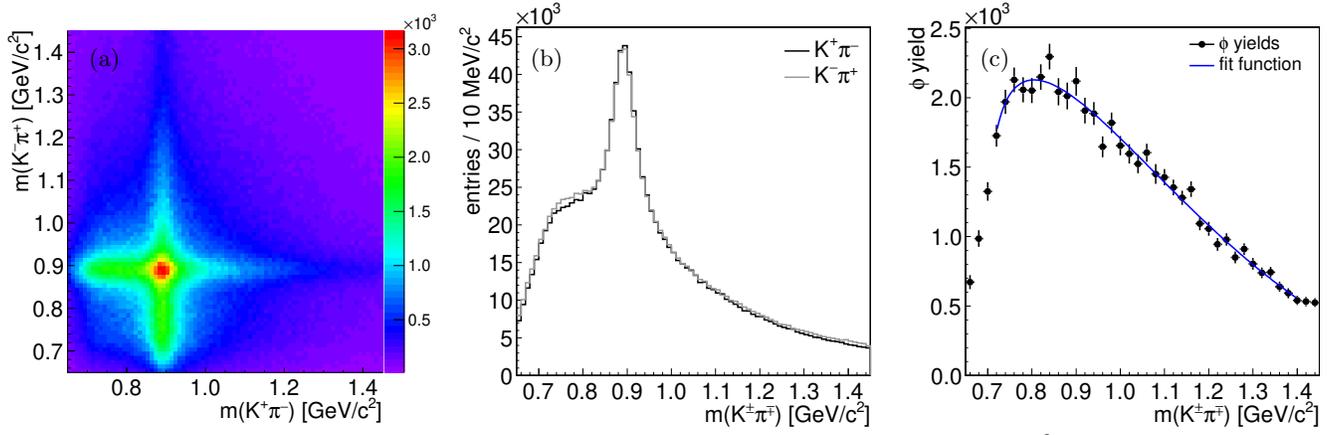
\paragraph{Supplemental Material  ---} 
There is a prominent $K^*(892)^0\bar{K^*}(892)^0$ contribution in real data (Fig.\,\ref{fig:SupplementaryMaterial}a). Since we are applying a slice-wise $\phi$-yield extraction method to determine the $K^+K^-\pi^+\pi^-$ event yield, we are not sensitive to (background) $K^{*0}\bar{K}^{*0} \to K^-K^+\pi^+\pi^-$ events. This has been validated by applying the $\phi$-yield extraction method in slices across the $m(K^\pm\pi^\mp)$ masses (Fig.\,\ref{fig:SupplementaryMaterial}a) for each charge combination individually (Fig.\,\ref{fig:SupplementaryMaterial}b). As demonstrated (Fig.\,\ref{fig:SupplementaryMaterial}c), no resultant $K^*(892)^0$ signal is visible, and we conclude that the method applied is not sensitive to such background. The enhancement at 1.8\,GeV/$c^2$ is thus not due to possible leakage of $K^*(892)^0$ events but are real $\phi\pi^+\pi^-$ events as expected to be produced close to the $K^*(892)^0\bar{K^*}(892)^0$ threshold via rescattering of the outgoing $K\bar{K}$ pair to form a $\phi(1020)$ and thus $\phi(1020)\pi^+\pi^-$ final state events.

\end{document}

%% file: authors_JS_Oct3rd_FN.tex
\affiliation{Polytechnic Sciences and Mathematics, School of Applied Sciences and Arts, Arizona State University, Tempe, Arizona 85287, USA}
\affiliation{Department of Physics, National and Kapodistrian University of Athens, 15771 Athens, Greece}
\affiliation{Ruhr-Universit\"{a}t-Bochum, Institut f\"{u}r Experimentalphysik, D-44801 Bochum, Germany}
\affiliation{Helmholtz-Institut f\"{u}r Strahlen- und Kernphysik Universit\"{a}t Bonn, D-53115 Bonn, Germany}
\affiliation{Department of Physics, Carnegie Mellon University, Pittsburgh, Pennsylvania 15213, USA}
\affiliation{Department of Physics, The Catholic University of America, Washington, D.C. 20064, USA}
\affiliation{School of Mathematics and Physics, China University of Geosciences, Wuhan 430074, People’s Republic of China}
\affiliation{Department of Physics, University of Connecticut, Storrs, Connecticut 06269, USA}
\affiliation{Department of Physics, Duke University, Durham, North Carolina 27708, USA}
\affiliation{Department of Physics, Florida International University, Miami, Florida 33199, USA}
\affiliation{Department of Physics, Florida State University, Tallahassee, Florida 32306, USA}
\affiliation{Department of Physics, The George Washington University, Washington, D.C. 20052, USA}
\affiliation{Physikalisches Institut, Justus-Liebig-Universit\"{a}t Gie{\ss}en, D-35390 Gie{\ss}en, Germany}
\affiliation{School of Physics and Astronomy, University of Glasgow, Glasgow G12 8QQ, United Kingdom}
\affiliation{Institut f\"ur Kernphysik, Goethe Universit\"at Frankfurt, D-60438 Frankfurt am Main, Germany}
\affiliation{GSI Helmholtzzentrum f\"{u}r Schwerionenforschung GmbH, D-64291 Darmstadt, Germany}
\affiliation{Helmholtz Forschungsakademie Hessen f\"ur FAIR (HFHF), Campus Frankfurt, GSI Helmholtzzentrum f\"{u}r Schwerionenforschung GmbH, D-64291 Darmstadt, Germany}
\affiliation{Institute of High Energy Physics, Beijing 100049, People's Republic of China}
\affiliation{Department of Physics, Indiana University, Bloomington, Indiana 47405, USA}
\affiliation{National Research Centre Kurchatov Institute, Moscow 123182, Russia}
\affiliation{Department of Physics, Lamar University, Beaumont, Texas 77710, USA}
\affiliation{Department of Physics, University of Massachusetts, Amherst, Massachusetts 01003, USA}
\affiliation{National Research Nuclear University Moscow Engineering Physics Institute, Moscow 115409, Russia}
\affiliation{Department of Physics, Mount Allison University, Sackville, New Brunswick E4L 1E6, Canada}
\affiliation{Department of Physics, Norfolk State University, Norfolk, Virginia 23504, USA}
\affiliation{Department of Physics, North Carolina A\&T State University, Greensboro, North Carolina 27411, USA}
\affiliation{Department of Physics and Physical Oceanography, University of North Carolina at Wilmington, Wilmington, North Carolina 28403, USA}
\affiliation{Department of Physics, Old Dominion University, Norfolk, Virginia 23529, USA}
\affiliation{Department of Physics, University of Regina, Regina, Saskatchewan S4S 0A2, Canada}
\affiliation{Department of Mathematics, Physics, and Computer Science, Springfield College, Springfield, Massachusetts, 01109, USA}
\affiliation{Thomas Jefferson National Accelerator Facility, Newport News, Virginia 23606, USA}
\affiliation{Laboratory of Particle Physics, Tomsk Polytechnic University, 634050 Tomsk, Russia}
\affiliation{Department of Physics, Tomsk State University, 634050 Tomsk, Russia}
\affiliation{Department of Physics and Astronomy, Union College, Schenectady, New York 12308, USA}
\affiliation{Department of Physics, Virginia Tech, Blacksburg, Virginia 24061, USA}
\affiliation{Department of Physics, Washington \& Jefferson College, Washington, Pennsylvania 15301, USA}
\affiliation{Department of Physics, William \& Mary, Williamsburg, Virginia 23185, USA}
\affiliation{School of Physics and Technology, Wuhan University, Wuhan, Hubei 430072, People's Republic of China}
\affiliation{A. I. Alikhanyan National Science Laboratory (Yerevan Physics Institute), 0036 Yerevan, Armenia}
\author{F.~Afzal\orcidlink{0000-0001-8063-6719}} \affiliation{Ruhr-Universit\"{a}t-Bochum, Institut f\"{u}r Experimentalphysik, D-44801 Bochum, Germany}
\author{M.~Albrecht\orcidlink{0000-0001-6180-4297}} \affiliation{Thomas Jefferson National Accelerator Facility, Newport News, Virginia 23606, USA}
\author{M.~Amaryan\orcidlink{0000-0002-5648-0256}} \affiliation{Department of Physics, Old Dominion University, Norfolk, Virginia 23529, USA}
\author{S.~Arrigo} \affiliation{Department of Physics, William \& Mary, Williamsburg, Virginia 23185, USA}
\author{V.~Arroyave} \affiliation{Department of Physics, Florida International University, Miami, Florida 33199, USA}
\author{A.~Asaturyan\orcidlink{0000-0002-8105-913X}} \affiliation{Thomas Jefferson National Accelerator Facility, Newport News, Virginia 23606, USA}
\author{A.~Austregesilo\orcidlink{0000-0002-9291-4429}} \affiliation{Thomas Jefferson National Accelerator Facility, Newport News, Virginia 23606, USA}
\author{Z.~Baldwin\orcidlink{0000-0002-8534-0922}} \affiliation{Department of Physics, Carnegie Mellon University, Pittsburgh, Pennsylvania 15213, USA}
\author{F.~Barbosa} \affiliation{Thomas Jefferson National Accelerator Facility, Newport News, Virginia 23606, USA}
\author{J.~Barlow\orcidlink{0000-0003-0865-0529}} \affiliation{Department of Physics, Florida State University, Tallahassee, Florida 32306, USA}\affiliation{Department of Mathematics, Physics, and Computer Science, Springfield College, Springfield, Massachusetts, 01109, USA}
% ...
\collaboration{The \textsc{GlueX} Collaboration}

%% file: acknowledgement.tex
\noindent
Acknowledgements

\noindent
We would like to acknowledge the outstanding efforts of the staff of the Accelerator and the Physics Divisions at Jefferson Lab that made the experiment possible. This work was supported in part by the U.S. Department of Energy, the U.S. National Science Foundation, the German Research Foundation, GSI Helmholtzzentrum f\"ur Schwerionenforschung GmbH, the Natural Sciences and Engineering Research Council of Canada, the Russian Foundation for Basic Research, the UK Science and Technology Facilities Council, the National Natural Science Foundation of China and the China Scholarship Council. This material is based upon work supported by the U.S. Department of Energy, Office of Science, Office of Nuclear Physics under contract DE-AC05-06OR23177. The work of F. Nerling is supported by the Helmholtz Forschungsakademie Hessen f\"ur FAIR. This research used resources of the National Energy Research Scientific Computing Center (NERSC), a U.S.
Department of Energy Office of Science User Facility operated under Contract No. DE-AC02-05CH11231. This work used
the Extreme Science and Engineering Discovery Environment (XSEDE), which is supported by National Science
Foundation grant number ACI-1548562. Specifically, it used the Bridges system, which is supported by NSF award
number ACI-1445606, at the Pittsburgh Supercomputing Center (PSC).

\vspace{0.3cm}